\documentclass[twocolumn,superscriptaddress,nobibnotes]{revtex4-1}

\usepackage{graphicx}
\usepackage{dcolumn}
\usepackage{bm}
\usepackage{times,mathptmx}
\usepackage{color}
\usepackage{multirow}
\usepackage{enumitem}
\usepackage{chngpage}
\usepackage{amsmath}
\usepackage[version=3]{mhchem}

\begin{document}	
\title{\bf Pockels Laser Directly Driving Ultrafast Optical Metrology}
\author{Shixin Xue}
\thanks{These two authors contributed equally.}
\affiliation{Department of Electrical and Computer Engineering, University of Rochester, Rochester, NY 14627}
\author{Mingxiao Li}
\thanks{These two authors contributed equally.}
\affiliation{Department of Electrical and Computer Engineering, University of California Santa Barbara, Santa Barbara, CA 93106}
\author{Raymond Lopez-rios}
\affiliation{Institute of Optics, University of Rochester, Rochester, NY 14627}
\author{Jingwei Ling}
\affiliation{Department of Electrical and Computer Engineering, University of Rochester, Rochester, NY 14627}
\author{Zhengdong Gao}
\affiliation{Department of Electrical and Computer Engineering, University of Rochester, Rochester, NY 14627}
\author{Qili Hu}
\affiliation{Institute of Optics, University of Rochester, Rochester, NY 14627}		
\author{Tian Qiu}
\affiliation{Institute of Optics, University of Rochester, Rochester, NY 14627}	

\author{Jeremy Staffa}
\affiliation{Institute of Optics, University of Rochester, Rochester, NY 14627}	
\author{Lin Chang}
\affiliation{Department of Electrical and Computer Engineering, University of California Santa Barbara, Santa Barbara, CA 93106}
\author{Heming Wang}
\affiliation{Department of Electrical and Computer Engineering, University of California Santa Barbara, Santa Barbara, CA 93106}
\author{Chao Xiang}
\affiliation{Department of Electrical and Computer Engineering, University of California Santa Barbara, Santa Barbara, CA 93106}

\author{John E. Bowers}
\email[]{e-mail: jbowers@ucsb.edu; qiang.lin@rochester.edu}
\affiliation{Department of Electrical and Computer Engineering, University of California Santa Barbara, Santa Barbara, CA 93106}
\author{Qiang Lin}
\email[]{e-mail: jbowers@ucsb.edu; qiang.lin@rochester.edu}
\affiliation{Department of Electrical and Computer Engineering, University of Rochester, Rochester, NY 14627}
\affiliation{Institute of Optics, University of Rochester, Rochester, NY 14627}



\begin{abstract}
\noindent\sffamily\textbf{Abstract}\linebreak
	\rmfamily 
 

 The invention of the laser unleashed the potential of optical metrology, leading to numerous advancements in modern science and technology. This reliance on lasers, however, also sets a bottleneck for precision optical metrology which is complicated by sophisticated photonic infrastructure required for delicate laser-wave control, leading to limited metrology performance and significant system complexity. Here we make a key step towards resolving this challenge, by demonstrating a Pockels laser with multi-functional capability that advances the optical metrology to a new level. The chip-scale laser exhibits a narrow intrinsic linewidth down to 167 Hz and a broad mode-hop-free tuning range up to 24~GHz. In particular, it offers an unprecedented frequency chirping rate up to 20~EHz/s, and an enormous modulation bandwidth $>$10~GHz, both orders of magnitude larger than any existing lasers. With this laser, we are able to successfully achieve velocimetry of 40~m/s at a short distance of 0.4~m, with a measurable velocity up to the first cosmic velocity at 1~m away, that is inaccessible by conventional ranging approaches, and distance metrology with a ranging resolution of $<$2~cm. Moreover, for the first time to the best of our knowledge, we are able to realize a dramatically simplified architecture for laser frequency stabilization, by direct locking the laser to an external reference gas cell without any extra external light control. We successfully achieve a long-term laser stability with a frequency fluctuation of only $\pm$6.5~MHz over 60 minutes. The demonstrated Pockels laser combines elegantly high laser coherence with ultrafast frequency reconfigurability and superior multifunctional capability that we envision to have profound impacts on many areas including communication, sensing, autonomous driving, quantum information processing, and beyond.

\end{abstract}
	
\maketitle
\noindent\sffamily\textbf{Introduction}\nolinebreak
	
\noindent\rmfamily

    Optical metrology has become one of the most effective ways for human to observe the world. By using light as the probe, it offers unprecedented advantages in measuring objects and physical quantities. In fundamental science, precision optical metrology plays indispensable roles in the most delicate experiments such as optical clock, gravitational wave detection, and dark matter observation \cite{ludlow2015optical,bailes2021gravitational,bertone2018new}. In daily life, the non-contact nature of optical measurement enables the capturing of target’s physical information in a fast and precise way, benefiting diverse applications ranging from self driving, robotics, to advanced manufacturing \cite{behroozpour2017lidar, li2020lidar, dahiya2009tactile, shimizu2021insight}. 

\begin{figure*}[htbp]
		\centering\includegraphics[width=2.0\columnwidth]{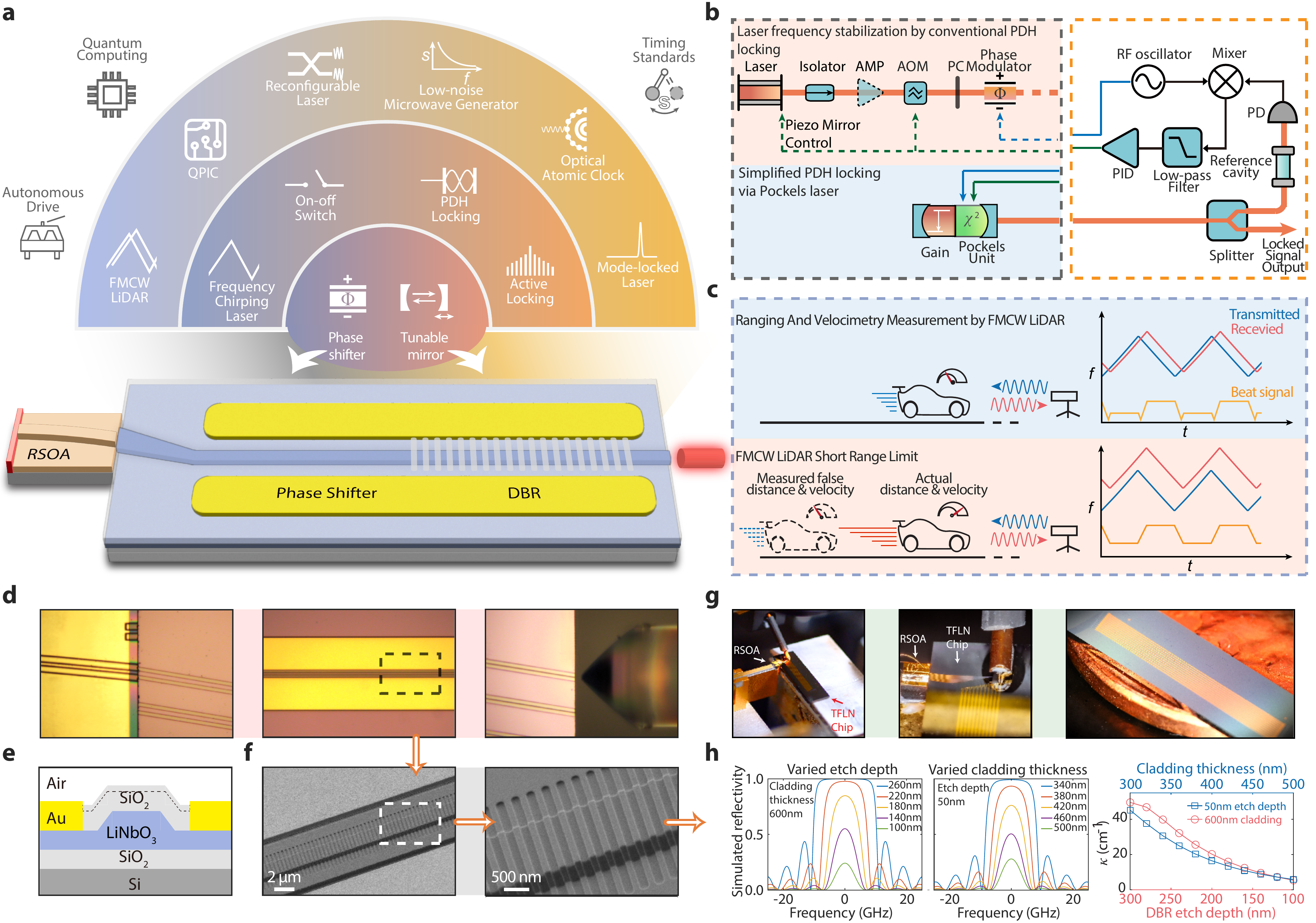}
		\caption{\label{Fig1} {\bf Concept of integrated Pockels laser for optical metrology applications.} {\bf (a)} Outlook of Pockels laser and its potential applications. Bottom: Schematic of the integrated eDBR-based Pockels laser. {\bf (b)} Schematic of laser frequency stabilization by the PDH locking. Top: Conventional PDH locking configuration. Bottom: Simplified PDH locking via Pockels laser. AOM: acousto-optic modulator; PC: polarization controller; AMP: optical amplifier; PD: photodetector. {\bf (c)} Schematic of distance metrology and velocimetry measurement by FMCW LiDAR. Top: normal detection regime for FMCW LiDAR. Bottom: false detection caused by large velocity and/or short distance. {\bf (d)} Optical microscopic images of the integrated eDBR-based Pockels laser. {\bf (e)} Schematic of the cross section of the DBR section. {\bf (f)} Scanning electron microscopic images of the DBR section. {\bf (g)} Left and Middle: Photo and its zoom-in of the laser device under testing. Right: Photo of the whole TFLN external cavity PIC chip (compared with a penny coin shown in the background). {\bf (h)} Simulated reflection spectrum (left two figures) and coupling coefficient $\kappa$ (right figure) of the DBR at varied etch depth and cladding thickness. The DBR grating in simulation has a period of 500~nm (corresponding to an optical frequency centered around 192.8~THz), with a total number of 20000 periods. }
	\end{figure*}

    The essential element lies in the heart of optical metrology is the generation and control of coherent laser waves. While significant advance has been made in recent years in the generation of high-coherence lasers particularly in narrow-linewidth semiconductor lasers \cite{liang2015ultralow,boller2020hybrid, xiang2021perspective, han2022recent, porter2023hybrid}, controlling laser waves for metrologic purposes remains fairly complicated. Realization of a complete metrologic functionality generally requires sophisticated photonic infrastructures external to the laser for manipulating the frequency, phase, power, polarization, and/or optical path of the laser wave. This complexity leads to bulky nature of current metrological systems which not only hinders the system miniaturization but also limits the metrology performance. Taking optical clock \cite{ludlow2015optical} as an example, lasers are required to stably lock to certain atomic transition lines and/or to a reference cavity for proper operation. It is generally realized via the Pound-Drever-Hall (PDH) technique \cite{drever1983laser, black2001introduction} that would require electro-optic components for frequency modulation, acoustic-optic devices for beam deflection, and piezoelectric (PZE) means for delicate servo control of the laser cavity \cite{alnis2008subhertz, schmid2019simple}, all of which can only be done external to the laser diode (Fig.~\ref{Fig1}(b)). The same complexity is inherited in an atomic quantum sensor \cite{fan2015atom, legaie2018sub} or an optical quantum computing system \cite{bruzewicz2019trapped, morgado2021quantum} whose operation also relies on precise laser accessing of atomic transitions.
    

On the other hand, in distance metrology and velocimetry, frequency-modulated continuous-wave (FMCW) light detection and ranging (LiDAR) technique has attracted significant interest recently given its simultaneous ranging and velocimetry capability and its superior environmental immunity, which is regarded as a crucial metrology technique supporting full self-driving in future motor vehicles \cite{behroozpour2017lidar, li2020lidar}. However, its dynamic range of ranging and velocimetry relies essentially on the speed and linearity of laser frequency chirping, which are significantly beyond the reach of current existing lasers. Consequently, FMCW LiDAR has to rely on complicated modulation and optoelectronic feedback control external to the laser for improving frequency chirping \cite{satyan2009precise, gao2012complex, lu2018broadband} whose limited performance seriously impacts the capability of velocity detection and could lead to severe false judgement in self-driving motor vehicles (Fig.~\ref{Fig1}(c)).

In this work, we make a key step towards resolving these challenges, by demonstrating a new type of laser that can directly drive optical metrology systems in a dramatically simplified configuration while with significantly enhanced performance. By utilizing our recently developed Pockels laser integration strategy and a novel co-tuned phase-distributed Bragg reflector (DBR) structure, we successfully combine narrow linewidth laser with unprecedentedly fast-speed, wide-range frequency reconfigurability. The chip-scale integrated laser exhibits a narrow intrinsic linewidth down to 167~Hz, a broad mode-hop-free (MHF) tuning range of 24~GHz, and a record high frequency chirping rate up to $2\times 10^{19}$~Hz/s. Moreover, the laser offers enormous modulation bandwidth up to $>$10~GHz for direct feedback locking inside the laser cavity, orders of magnitude larger than any existing lasers. 

The superior performance of the demonstrated laser now opens up a great opportunity to significantly advance a variety of optical metrologic applications. To showcase this capability, we use the laser to realize an FMCW LiDAR and achieve ultrafast ranging and velocimetry of distant objects, with a measured velocity up to 40~m/s at a very short distance of 0.4 meter that is inaccessible in current state-of-the-art FMCW LiDAR. At the same time, We also achieve a two-dimensional imaging with a ranging resolution of $<$2.0~cm. 
 On the other hand, we successfully demonstrate a dramatically simplified laser-frequency stabilization architecture, by direct feedback locking the laser to a reference gas cell without any external light control/modulation. We are able to achieve a long-term laser frequency stability of $\pm$6.5~MHz over 60 minutes. 

    \begin{figure*}[t!]
		\centering\includegraphics[width=2.0\columnwidth]{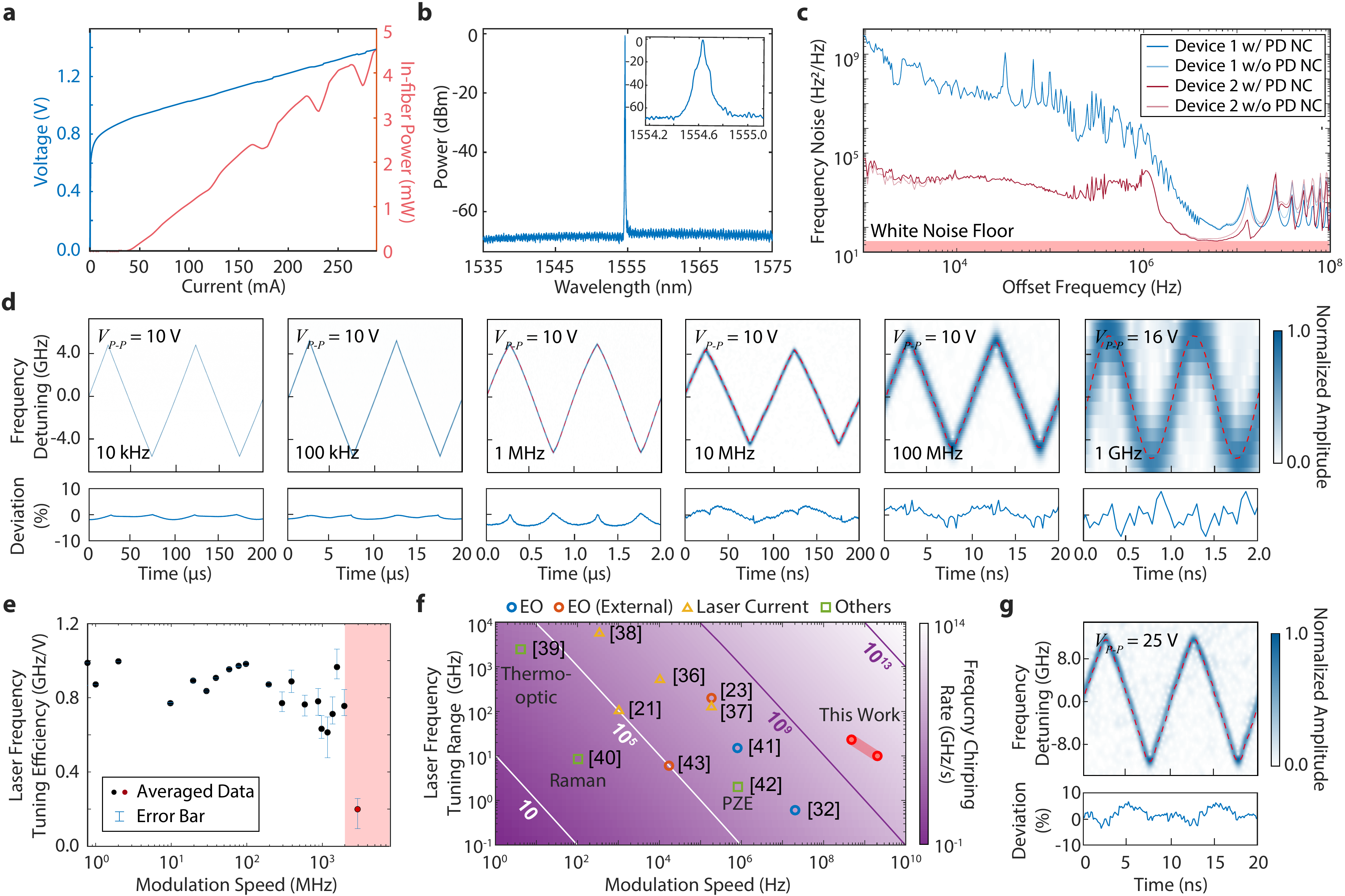}
		\caption{\label{Fig2}  {\bf Characterization of laser performance.} {\bf (a)} Recorded LIV curve of the laser, shown as laser power recorded in fiber. The corresponding laser output power on chip is 5~dB higher, determined by the chip-to-fiber coupling efficiency. {\bf (b)} Optical spectrum of single mode lasing. The inset shows the zoom-in spectrum. {\bf (c)} Recorded frequency noise spectrum of two laser devices, with the white noise floor highlighted in shaded red. PD NC: photodetector noise cancellation via cross-correlation technique. {\bf (d)} Recorded time-frequency spectrograms of the beat note between the Pockels laser and a reference diode laser at different modulation frequencies. The lower panels show the corresponding linearity deviation of laser frequency chirping. The red dashed line in the figures shows the temporal waveform of the driving electrical signal, which is produced by an arbitrary waveform generator (AWG). For modulation speed $\ge$~1~MHz, broadband electric amplifiers are used to boost the driving signal before it is applied to the laser. The limited bandwidths of the electric amplifiers lead to certain waveform distortion on the driving electric signal, particularly at high modulation frequencies. $V_{p-p}$ shown on the figure indicates the peak-to-peak voltage of the driving electric signal. Details in Supplementary Information. {\bf (e)} Recorded laser-frequency tuning efficiency as a function of modulation speed ($\emph{i.e.}$, modulation frequency). The red shaded area indicates the frequency region where the tuning efficiency degrades. The error bars indicate the processing uncertainty of short-time Fourier transform caused by the instrument limitation. {\bf (f)} Comparison of state-of-the-art laser frequency tuning performance. Red dots highlight the results of our devices. {\bf (g)} Same as {\bf (d)}, at a modulation speed of 100~MHz but with a \textit{V}$_{p-p}$ of 25 Volts.}
    \end{figure*}
    
\noindent\sffamily\textbf{Results}\nolinebreak
	
\noindent\rmfamily\textbf{Laser design and basic performance.}

The demonstrated laser is based upon hybrid integration between an InP reflective semiconductor optical amplifier (RSOA) gain chip and a thin-film lithium niobate (TFLN) photonic integrated circuit (PIC) functioning as an external cavity of the laser (Fig.~\ref{Fig1}(a)). TFLN PIC has attracted significant interest recently for broad applications \cite{lin2020advances, zhu2021integrated, boes2023lithium}. Using its electro-optic effect and quadratic optical nonlinearity for laser operation results in a novel type of integrated laser, namely integrated Pockels laser \cite{li2022integrated}, that opens up great potential for broad applications in a wide variety of photonic functionalities (Fig.~\ref{Fig1}(a)). However, integrated lasers previously demonstrated on the TFLN platform \cite{li2022integrated, op2021iii, gao2021chip, shams2022electrically, han2022widely, snigirev2023ultrafast, li2023high, luo2023advances} suffer from a fairly large laser linewidth and limited frequency chirping range, unsuitable for optical metrology applications. 

To resolve this challenge, we employ the extended DBR (eDBR) approach \cite{huang2019high} for single-mode lasing. In particular, we develop a novel type of eDBR structure (Fig.~\ref{Fig1}(e,f)) in which the Bragg grating is defined in the low-index silicon-oxide cladding layer rather than in the TFLN layer itself. This novel approach allows for flexible engineering the strength of Bragg scattering simply by controlling the layer thickness of the silicon oxide cladding and/or the etching depth of the Bragg grating (Fig.~\ref{Fig1}(e),(h)). As silicon oxide has a refractive index considerably lower than TFLN while its plasma etching quality is significantly higher, this approach allows to realize ultra-low Bragg scattering strength down to \(\kappa\) = 5.4 cm$^{-1}$ over a long grating length of 1 cm, resulting in an ultra-narrow reflection spectrum as shown in Fig.~\ref{Fig1}(h), ideal for narrow-linewidth lasing. For high-speed tuning the laser frequency, a pair of tuning electrodes are integrated with the eDBR to tune its center wavelength, by use of the electro-optic Pockels effect ($r_{33}$) of TFLN. For MHF tuning the laser frequency, the driving electrodes are extended to a section of TFLN waveguide which functions effectively as a phase shifter to assist the tuning of the resonance mode of the whole laser cavity (Fig.~\ref{Fig1}(a)). This co-tuned approach simplifies the electric driving structure and laser operation. Compared with ring-resonator-based laser structures, the eDBR Pockels laser design mitigates the mode-mismatching issues resulting from fabrication error and enables plug-and-play functionality, facilitating a more versatile laser tuning. Its potential functionalities are thus expanded as we will show below, surpassing conventional laser approaches. Details of the laser structure design is provided in the Method.

        \begin{figure*}[t!]
		\centering\includegraphics[width=2.0\columnwidth]{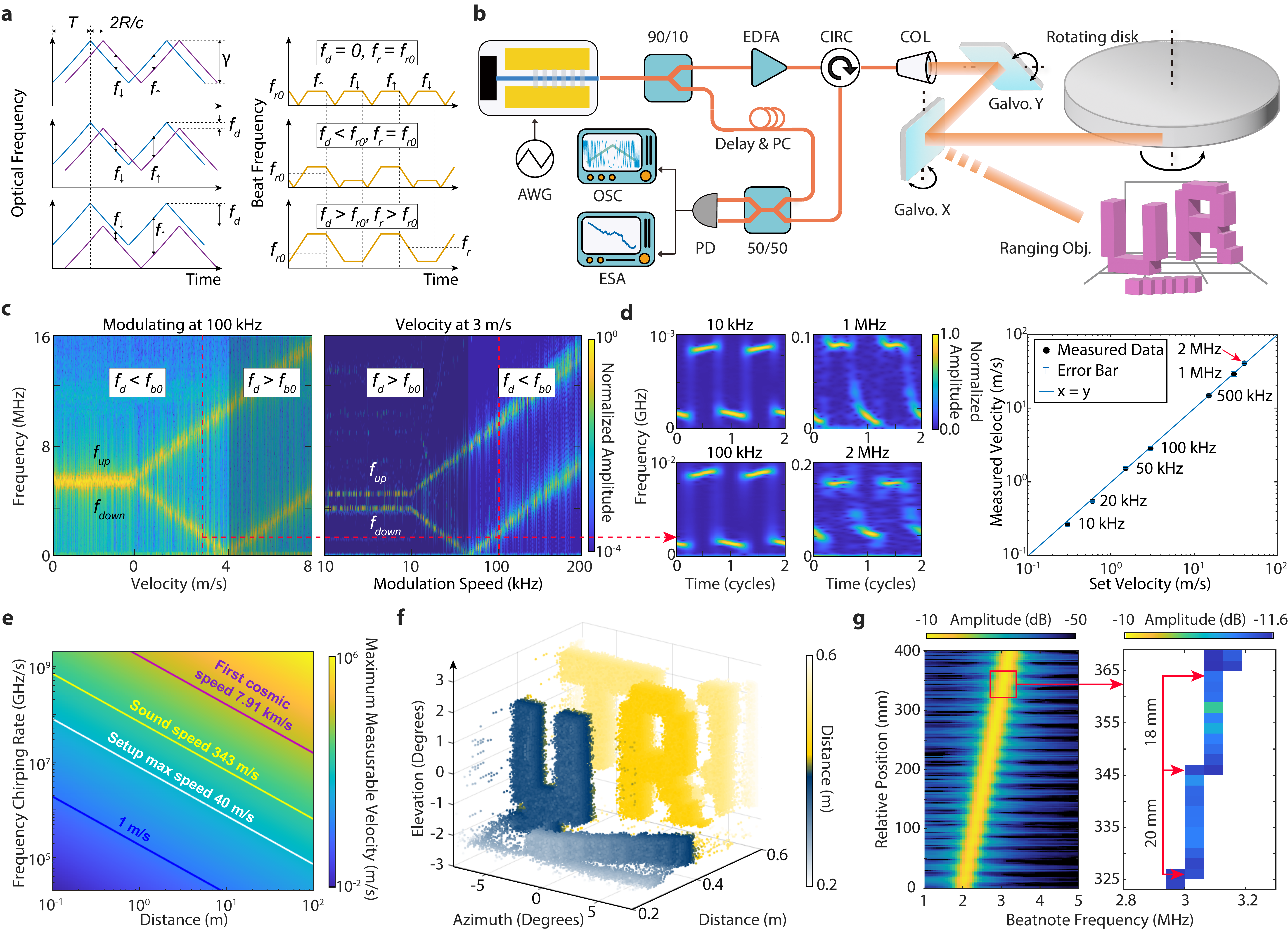}
		\caption{\label{Fig3} {\bf Ranging and velocimetry performance.}  {\bf (a)} Schematic illustrating the time-dependent optical (left) and beatnote (right) frequency in an FMCW LiDAR. Top, middle, and bottom panels show three operation regimes with $f_d =0$, $f_d < f_r$, and $f_d > f_r$, respectively. {\bf (b)} Experimental setup for distance and velocity measurement. Photos of the objects are provided in the SI. EDFA: erbium-dope fiber amplifier; CIRC: circulator; COL: collimator; Galvo.: Galvo mirror; PC: polarization controller; PD: photodetector; OSC: oscilloscope; ESA: electric spectrum analyzer. {\bf (c)} Left: Spectrum of the recorded beating signal as a function of target velocity, with a fixed laser frequency modulation speed of 100~kHz. Right: That as a function of laser frequency modulation speed, at a fixed target velocity of 3~m/s. The shaded areas indicate the regime of false detection caused by inadequate laser frequency chirping rate. The red dashed line highlights the case with a modulation speed of 100~kHz and a target velocity of 3~m/s, whose detailed spectrogram is shown in {\bf (d)}. {\bf (d)} Time-frequency spectrogram of the beating signal at four different target velocities of 0.3, 3, 30, and 40~m/s. To showcase the effect of laser frequency modulation speed, the four spectrograms are measured with different laser-frequency modulation speed of 10~kHz, 100~kHz, 1~MHz, and 2~MHz, respectively. The right figure compared the measured velocity by the device with the set velocity of the object. {\bf (e)} Expected maximum velocity that can be measured by our laser. {\bf (f)} 2D scanned ranging image of a static object. {\bf (g)} Spectrum of the beatnote as a function of relative position of an object, with a step of 2~mm. The red area is highlighted at right to show the ranging resolution.}
	\end{figure*}

With this approach, we are able to achieve single-mode lasing operation with very narrow lasing linewidth. Figure \ref{Fig2} shows the laser performance. The laser exhibits a threshold current of 45~mA and emits an optical power of about 13~mW on chip (corresponding to 4.1~mW recorded in fiber) with a driving current of 260~mA (~Fig.~\ref{Fig2}(a)). The laser operates around a telecom wavelength of 1555~nm, with a side-mode suppression ratio $>58$~dB (Fig.~\ref{Fig2}(b)). In particular, the laser exhibits very narrow linewidth as shown in Fig.~\ref{Fig2}(c), with a white frequency noise floor down to 26.7 Hz$^2$/Hz which corresponds to an intrinsic linewidth of only 167~Hz, more than an order of magnitude smaller than others demonstrated on the TFLN platform  \cite{li2022integrated, op2021iii, gao2021chip, shams2022electrically, han2022widely, snigirev2023ultrafast, li2023high, luo2023advances}. More details and discussion of the device linewidth are included in the Methods and SI. 

High-speed frequency modulation of a laser is essential for a variety of optical metrology applications. To show this capability, we apply an electric signal with a triangular waveform to drive the TFLN external laser cavity, and monitor the laser frequency tuning by heterodyning the laser output with a stable narrow-linewidth continuous-wave reference laser. Figure \ref{Fig2}(d) shows the recorded waveforms of the beating frequency at different modulation frequencies. It shows clearly that the laser frequency tuning follows faithfully the driving electric signal at all modulation frequencies up to 1~GHz. The blurring of the recorded spectrogram at a high modulation speed above 100~MHz is simply due to the limited sampling rate of the real-time oscilloscope used for recording the beating signal. The frequency chirping nonlinearity is less than 1\% (root-mean square (RMS) value) at modulation frequencies of 10 and 100~kHz. It slightly increases to 2.2\%-3.8\% for higher modulation frequencies, which is likely caused by the electric amplifier used to boost the driving electric signal (see the caption of Fig.~\ref{Fig2}(d)). A MHF tuning range up to 10~GHz is achieved at all these modulation frequencies. The frequency tuning efficiency maintains roughly $\sim$0.8~GHz/V for modulation speed up to 2~GHz (Fig.~\ref{Fig2}(e)) after which it starts to drop. The frequency tuning range of 10~GHz at a modulation speed of 1~GHz corresponds to a frequency chirping rate as high as $2 \times 10^{19}$~Hz/s. Figure \ref{Fig2}(f) compares the frequency tuning performance of the laser with state--of-the-art \cite{satyan2009precise,vasilyev2012terahertz, behroozpour2016electronic, dilazaro2018large,zheng2022high,li2024microcavity, zhang2024high, snigirev2023ultrafast, lihachev2022low,lu2018broadband,wang2018inverse}. It shows clearly that the recorded performance here is the highest among any existing lasers in terms of both frequency chirping rate and modulation speed. Interestingly, the MHF frequency tuning range increases considerably to 24~GHz at a modulation speed around 100~MHz (Fig.~\ref{Fig2}(g)), which is more than one order of magnitude larger than previously reported \cite{li2022integrated, snigirev2023ultrafast, li2023high}. It is likely due to the improved coordinated electro-optic tuning between the phase shifter and the eDBR around this frequency region (see Supplementary Information (SI) for detailed discussion of coordinated tuning). 



  \begin{figure*}[htbp]
		\centering\includegraphics[width=2.0\columnwidth]{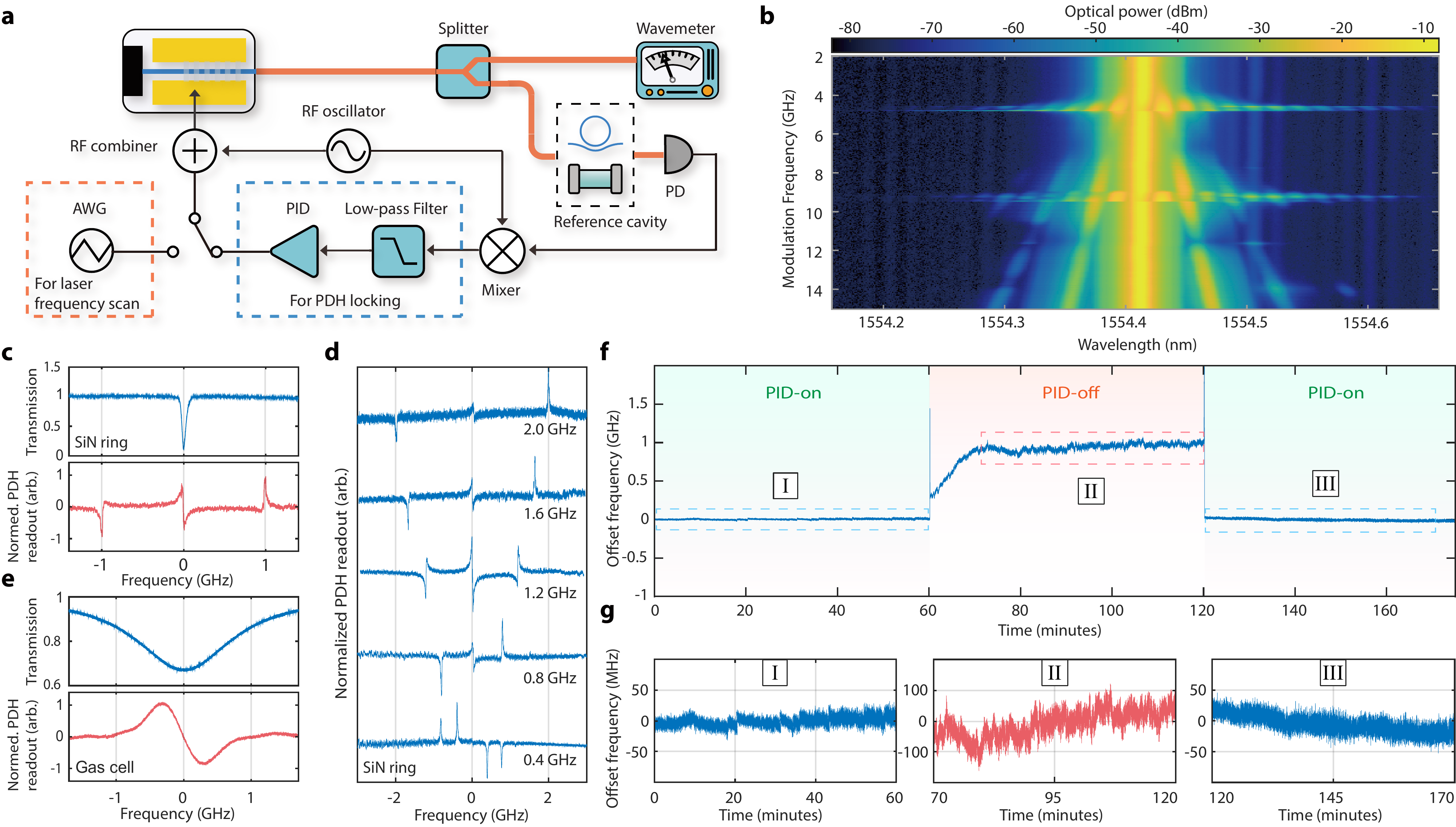}
    \caption{\label{Fig4}  {\bf Laser frequency stabilization.} {\bf (a)} Schematic of experimental setup for PDH laser locking. The reference cavity is either a high-Q silicon intride microring resonator or a H$^{13}$C$^{14}$N gas cell. To obtain laser-scanned transmission spectrum of the reference cavity ({\bf (c), (e)}, Top figures), a triangular-waveformed signal, produced by an AWG at a modulation speed of 200~Hz, is used to drive the TFLN laser external cavity, so as to linearly scan the laser frequency at a slow speed. To obtain the laser-scanned PDH error signal ({\bf (c), (e)}, Bottom figures, and {\bf (d)}), A single-frequency sinusoidal RF signal is added to drive the laser external cavity, together with the triangular-waveformed signal. For feedback locking the laser, the AWG is switched off and the PDH error signal is directed to a PID servo whose output is used to drive the laser external cavity.  {\bf (b)} Optical spectrum of laser output when laser external cavity is modulated by a sinusoidal RF signal at different frequencies, with an RF power of 7 dBm.  {\bf (c)} Laser-scanned transmission spectrum (top) and PDH error signal (bottom) of a resonance mode of the reference silicon nitride microresonator. {\bf (d)} Laser-scanned PDH error signal of the silicon nitride resonator, at different RF modulation frequencies. {\bf (e)} Laser-scanned absorption spectrum (top) and PDH error signal (bottom) of the H$^{13}$C$^{14}$N P16 line of the reference gas cell. A RF modulation speed of 630~MHz is used to obtain the figure of PDH error signal. {\bf (f)} Time-dependent laser frequency, measured by a wavemeter ((Toptica Photonics, WS6-600 VIS/IR-I)). The PID servo is switched on in the first and third hour, and is switched off in the second hour. {\bf (g)} Detailed time-dependent laser frequency at the three time sections in dashed area in (f).}
\end{figure*}

\medskip
 
\noindent\rmfamily\textbf{Distance metrology and velocimetry}

The superior linear frequency tuning performance of the laser enables FMCW LiDAR with not only high-resolution ranging but particularly also velocimetry with a very large dynamic range. FMCW LiDAR relies on detecting the beating frequency between the reflected and reference laser waves for retrieving the distance ($R$) and velocity ($v$) information. The two beating frequencies during the laser frequency ramping-up and ramping-down duration are given by  \cite{behroozpour2017lidar, li2020lidar}
\begin{eqnarray}
f_{\uparrow} &=& \gamma \frac{2R}{c} - \nu_o \frac{2v}{c} \equiv f_r - f_d, \label{f_up}\\
f_{\downarrow} &=& \gamma \frac{2R}{c} + \nu_o \frac{2v}{c} \equiv f_r + f_d, \label{f_down}
\end{eqnarray}
where $\gamma$ is the frequency chirping rate, $\nu_o$ is the center laser frequency, and $c$ is the velocity of light in vacuum. The first term $f_r \equiv \gamma \frac{2R}{c}$ is the frequency difference introduced by time delay due to light propagation, and the second term $f_d \equiv \nu_o \frac{2v}{c}$ represents the frequency shift induced by the Doppler effect. The detected distance and velocity are thus given by 
\begin{eqnarray}
R = \frac{c}{4\gamma}(f_{\uparrow} + f_{\downarrow}), \qquad v = \frac{c}{4\nu_o}(f_{\downarrow} - f_{\uparrow}). \label{Rv}
\end{eqnarray}
Equations (\ref{f_up})-(\ref{Rv}) show that meaningful distance and velocity measurement requires both $f_\uparrow > 0$ and $f_\downarrow > 0$ which in turn requires $|f_d| < |f_r|$, leading to $|v| < \frac{\gamma R}{\nu_o}$. The detectable velocity is fundamentally limited by the laser frequency chirping rate and the ranging distance. If $|f_d| > |f_r|$, FMCW LiDAR will produce false values of both distance and velocity. Figure \ref{Fig3}(a) illustrates this effect. In practice, $f_\uparrow$ and $f_\downarrow$ are required to be $>~ \sim 1$~kHz to reduce the impact of the $1/f$-noise of optical detector, which further limits the dynamic range of velocimetry. In general, it is challenging for a conventional FMCW LiDAR to detect a fast moving object at a short distance. 


This challenge is elegantly resolved by our laser. To show this feature, we use the setup shown in Fig.~\ref{Fig3}(b) to perform velocimetry on a high-speed target, which is an 8-inch foam disk mounted on a high-speed DC motor with tunable rotation speed (See SI for a photo). The laser beam is aligned with the edge of the disk to capture the high tangential velocity. The laser frequency is modulated with a triangular waveform with a frequency chirping range of 10~GHz but at different modulation speeds to test the ranging and velocimetry performance. 

First, we fix the modulation speed at 100~kHz (which corresponds to $\gamma = 2$~PHz/s) and change the target velocity between 0 and 8~m/s. Figure~\ref{Fig3}(c) (left) shows the recorded spectra of the beating signal. When $v=0$, $f_\uparrow = f_\downarrow = 5.31$~MHz which indicates an effective target distance of 0.4~m. With increased targeted velocity, $f_\downarrow$ increases linearly with $v$ while $f_\uparrow$ decreases linearly, as expected, till $v=$ 4~m/s at which $f_\uparrow $ approaches zero. With further increased $v$, both $f_\uparrow$ and $f_\downarrow$ increase with $v$ since in this region, Doppler frequency shift dominates over the time-delay-induced frequency difference, $|f_d| > |f_r|$. Consequently, the LiDAR will infer an incorrect distance longer than the real value and an incorrect velocity smaller than the real value (see Eq.~(\ref{Rv})). This effect can be seen more clearly by fixing the target velocity at $v=$3~m/s while changing the laser-frequency modulation speed from 10 to 200~kHz (corresponding to $\gamma = 0.2-4$~PHz/s), as shown in Fig.~\ref{Fig3}(c)(right). When the modulation speed is low, $f_\uparrow$ erroneously decreases with increased modulation speed till approaching zero, after which it starts to increase linearly. Again, it shows clearly that the LiDAR will produce incorrect ranging and velocimetry information with an insufficient rate of laser frequency chirping.

To show the capability for detecting large velocity, we increase the target velocity up to $\sim$40~m/s. Figure \ref{Fig3}(d) show the spectrograms of the recorded beating signal at different target velocities. They show clearly that both $f_\uparrow$ and $f_\downarrow$ can be well resolved at all different velocities, with adequate laser frequency chirping rate. The waveform distortion of the time-dependent beatnote frequency is primarily introduced by the nonlinearity of laser frequency chirping which is in turn dominated by the waveform distortion of the driving electric signal (See Fig.~\ref{Fig2}(d) and caption). Such waveform distortion is generally the dominant factor determining the precision of velocity measurement. For example, at the target velocity of 40~m/s, the recorded $f_d = (f_\downarrow - f_\uparrow)/2 = 52.7 \pm 2.0$~MHz, which corresponds to a measured velocity of $41.0 \pm 1.5$~m/s. Overall, the measured velocities agree well with the set values over the entire velocity range of 0.3$-$40~m/s. 40~m/s corresponds to 89.5 miles/hour, which covers pretty much the majority range of vehicle velocity on the road if the LiDAR is used for self driving. Note that, although we use different frequency chirping rates in Fig.~\ref{Fig3}(d) to showcase the chirping rates required for detecting different velocities, in practice, $\gamma$ can be fixed at a certain high value for a desired dynamic range of velocity.

As 40~m/s is the maximum velocity offered by the rotation disk, we cannot measure larger velocity values in the experiment. However, the high laser frequency chirping rate implies that much larger velocity could be measured by the laser. Figure \ref{Fig3}(e) shows the expected measurable velocity by our laser at different chirping rates and ranging distances. Velocity up to the first cosmic speed of 7.91~km/s could be measured by our laser at a distance of 1 meter away, and velocity up to $\sim$1000~km/s could be measured for an object about 100 meters away. This clearly shows the powerfulness of the demonstrated laser for velocimetry with an extremely large dynamic range. 

In addition to the velocimetry capability shown above, the laser also offers high ranging resolution. To show this, The collimated laser beam is directed by the Galvo mirrors on another path to scan across a different object. The static object consists of two letters and a background board, with each separated by a distance of 8~cm. Moreover, a series of blocks with a 1-cm depth step is placed at the bottom to illustrate the depth resolution (see Fig.~\ref{Fig3}(b). A photo of the object is provided in SI). The laser-frequency modulation speed is set to 60 kHz for easier data acquisition. Figure \ref{Fig3}(f) shows the recorded two-dimensional image of the ranging object whose detailed structures are clearly observed. Detailed calibration of distance measurement (Fig.~\ref{Fig3}(g)) shows a ranging resolution of $<$2~cm, consistent with the laser frequency chirping range of 10~GHz (which corresponds to a theoretical ranging resolution of 1.5~cm).


\medskip

\noindent\rmfamily\textbf{Laser frequency stabilization}

In addition to the fast linear frequency chirping, the Pockels laser also supports high-speed phase modulation directly inside the laser cavity. This is shown in Fig.4(b) in which a sinusoidal radio frequency (RF) driving signal is now applied to drive the TFLN external laser cavity, with the modulation speed varying between 2 to 15~GHz. Created modulation sidebands are clearly visible on the optical spectrum of the laser output. At a specific modulation speed of around 4.7~GHz and 9.4~GHz that match the free-spectral range of the laser cavity, the RF modulation excites active fundamental and harmonic mode locking which significantly broadens the optical spectrum of the laser. Other than that, the created modulation sidebands evolve smoothly with increased modulation speed over the entire range of modulation speed up to 15~GHz. To verify the nature of laser modulation, we remove the RSOA gain chip, and launch a continuous-wave laser to the passive TFLN external cavity device and monitor the transmission spectrum during the same modulation. The results are provided in SI. The similarity between the two figures clearly verifies the nature of direct phase modulation inside the laser cavity.


The high-speed phase modulation feature of the laser opens up a great avenue for laser frequency stabilization with dramatically simplified architecture, as the embedded phase shifter eliminates the need for additional bulky external modulator and other alignment-sensitive components. To show this feature, we first use a high-Q silicon nitride microring resonator (intrinsic optical Q of 9 million) as a reference cavity, and use the PDH approach to monitor the laser modulation output. To do so, we apply a triangular-waveform driving signal with a slow modulation speed of 200 Hz to scan the laser frequency. At the same time, we apply a sinusoidal RF driving signal to create modulation sidebands. Figure 4(d) shows the recorded PDH error signal at different RF modulation frequencies from 0.4 to 2.0~GHz, which shows clear feature of PDH error signal.

The clean PDH error signal directly produced by the laser indicates the great potential for direct PDH locking of the laser frequency. To show this, we switch the reference cavity to a fiber-coupled H$^{13}$C$^{14}$N gas cell. The laser wavelength is adjusted to be around the P16 absorption line near 1554.59 nm. Figure~\ref{Fig3}(e) shows the recorded laser-scanned spectra of P16 absorption line of H$^{13}$C$^{14}$N and the corresponding PDH error signal. To lock the laser frequency, we remove the triangular-waveform RF signal and direct the produced PDH error signal to a proportional-integral-derivative (PID) servo controller whose output is combined with the sinusoidal RF signal to directly drive the TFLN external laser cavity. We record the laser frequency with a wavemeter, which is shown in Fig.~\ref{Fig3}(f). 

To show the repetitive direct locking of laser frequency, we turn the PID servo on for 60 minutes (Duration I in Fig.~\ref{Fig4}(f)), and then switch it off for another 60 minutes (Duration II in Fig.~\ref{Fig4}(f)) during which the lases is free running. After that, we turn the PID servo on again for another 60 minutes (Duration III in Fig.~\ref{Fig4}(f)). As shown in Duration III, the laser frequency is stabilized to a relatively constant value when the PID servo is on, with a frequency fluctuation of only $\pm 13.1$~MHz (RMS value) (Section III in Fig.~\ref{Fig4}(g)). This value is actually limited by the resolution of the wavemeter used (Toptica Photonics, WS6-600 VIS/IR-I) which is about 20~MHz. The actual laser frequency fluctuation could be smaller, which will require further characterization in the future when a better wavemeter becomes available. A small laser frequency drift of about 42.4~MHz is observed over a time duration of 50 minutes, which is simply due to the mechanical drift of the two stages holding the RSOA and TFLN chips (Fig.~\ref{Fig1}(g)). To verify this, we manually correct the mechanical drift in Duration I, by 50~nm every about 10 minutes. As shown in the section I of Fig.~\ref{Fig4}(g), the laser frequency is now fully stabilized to a constant value over the entire 60 minutes, with a frequency fluctuation of about $\pm 6.5$~MHz (RMS value) similar to that of Duration III. When the PID servo is turned off (Duration II), the laser frequency starts to drift as the laser is free running. The laser frequency drifts by about 600~MHz within 10 minutes and then settles to a relatively stable value after which the laser frequency drifts slowly with an amount of $\sim$113~MHz over a time duration of 50 minutes. Also, the laser frequency fluctuations increases to $\pm$44.5~MHz (RMS value) (Section II in Fig.4(g)). These observations clearly verify the powerfulness of the demonstrated approach for direct laser frequency locking.

\medskip

\noindent\rmfamily\textbf{Discussion}

The frequency stability of the current laser is primarily determined by the mechanical stability of the testing station since the RSOA gain chip and the TFLN PIC chip are placed on separate stages ({Fig.~\ref{Fig1}(g)}). We expect it can be improved significantly by using an appropriate packaging approach \cite{shen2020integrated} in the future, which would also further improve the performance of laser frequency stabilization. On the other hand, in the current laser, we use one set of driving electrode to co-tune the phase shifter and the eDBR. Although it helps simplify the laser operation, it limits the MHF tuning range since large MHF tuning of laser frequency requires coordinated tuning of the phase shifter and the eDBR (see SI for details). We expect the MHF tuning range can be increased considerably with optimized driving electrodes for independent tuning the phase shifter and the eDBR. 

In summary, we have demonstrated a chip-scale integrated Pockels laser in the telecom band that exhibits a narrow intrinsic linewidth down to 167 Hz, emits an optical power of 13~mW on chip, and supports a broad MHF tuning range of 24~GHz. In particular, the laser offers an unprecedented frequency chirping rate up to 20~EHz/s, and an enormous modulation bandwidth up to $>$10~GHz, both orders of magnitude larger than any existing lasers. With this laser, we are able to successfully achieve velocimetry of 40~m/s at a short distance of 0.4~m that is inaccessible by conventional FMCW LiDAR, and distance metrology with a ranging resolution of $<$2~cm. Moreover, for the first time to the best of our knowledge, we are able to realize a dramatically simplified architecture for laser frequency stabilization, by direct locking the laser to an external reference gas cell without any extra external light control. We successfully achieved a long-term laser stability with a frequency fluctuation of only $\pm$6.5~MHz over 60 minutes. 

The outstanding performance of the demonstrated laser is expected to have profound impacts on a wide range of optical metrology applications. For example, in optical clocks, atomic sensors, and optical quantum computing systems, significant advances have been made recently in miniaturization of optical reference cavity \cite{lee2013spiral, zhao2021integrated}, atomic vapor cell \cite{kitching2018chip}, and ion trap \cite{bruzewicz2019trapped}. However, development of laser frequency control remains fairly limited. To date, lasers in these applications functions pretty much solely as light sources whose control requires complex and bulky systems as discussed in the Introduction. This becomes a major challenge limiting the size, weight, and power consumption, hindering their wide deployment in practical environment. As a majority of optical metrology applications require frequency-stabilized lasers for proper operation, the simplified architecture demonstrated here would significantly advance the integration and performance of optical metrology systems. As another example, photonic Doppler velocimetry \cite{dolan2020extreme} plays a crucial role in the detection of shock waves and the study of dynamic compression, which is indispensable for applications such as inertial confinement fusion, explosive detonation detection, etc. However, detection of ultrahigh velocity remains challenging since the resulting extremely large Doppler frequency shift imposes serious burden on the operation bandwidth of optical detectors. This challenge could potentially be resolved with our laser whose ultrafast frequency chirping can produce a delay-induced frequency to offset the Doppler frequency shift (see Eqs.~(\ref{f_up}) and (\ref{f_down})). Beside optical metrology applications, the demonstrated Pockels laser combines elegantly high laser coherence with ultrafast frequency reconfigurability and superior multi-functional capability that we envision to be of great promise for a wide range of applications including communication, sensing, optical and microwave synthesis, among many others.

\medskip
\newpage
\noindent\rmfamily\textbf{Method}

\medskip
\noindent\rmfamily\textbf{Laser structure design:}
The extended-DBR laser structure contains a gain section, a passive section connecting gain and EO-tunable component, a phase shifter section, and a tunable eDBR section. As shown in the schematic from {Fig.~\ref{Fig1}(a)}, a hybrid integration approach is implemented for the laser: a single-angled facet (SAF) III/V gain chip operating in the C-band is edge-coupled to an external cavity (EC) made on a TFLN PIC chip. A 5-\textmu m-wide horn coupler at the EC facet is optimized to reduce mode mismatch between the III-V gain chip and the LN waveguide, minimizing the intracavity losses. An angle of 10 degrees is designed for the waveguide input facet to reduce back reflection of light. Laser frequency control is achieved by positioning a pair of electrodes along a 3-mm-long eDBR section and a 6-mm-long TFLN waveguide, the latter of which functions effectively as an electro-optic (EO) phase shifter. A 3.0-\textmu m gap between the waveguide and the gold electrode is designed to optimize the EO tuning efficiency while keeping the propagation loss intact. \textit{V}$_{\pi}$ of the EO phase shifter is measured to be about 6 V, which is characterized with a separate device with an identical structure. Details of the length of each section to optimize the mode-hop-free tuning range can be found in the Supplementary Information.

\medskip
\noindent\rmfamily\textbf{Device fabrication:}
The devices were fabricated on a 600-nm-thick x-cut single-crystalline LN thin film bonded on a 4.7-${\rm \mu}$m silicon dioxide layer sitting on a silicon substrate (from NanoLN). Fabrication began with waveguide patterning: a layer of ZEP520A resist was spun onto the sample and baked, followed by electron beam lithography (EBL) to define the waveguide structure. The sample then underwent argon ion (\ce{Ar^{+}}) dry etching to form ridge waveguides with an etch depth of 300 nm. A subsequent cleaning process removed the resist and redeposition residues, resulting in smooth waveguide sidewalls. To define the eDBR grating in the demonstrated device, a 500nm-thick \ce{SiO2} cladding was deposited using ICP-PECVD. A second EBL step was employed to pattern the DBR structure, followed by reactive ion etching (RIE) to etch down 150 nm, creating the grating trenches. For electrodes fabrication, we first etched the \ce{SiO2} down to the LN layer using another lithography and etching step. Gold electrodes (500 nm thick) were then deposited by electron-beam evaporation, followed by a PMMA liftoff process. The chip facets are cut and polished to have a smooth coupling with the gain chip.

\medskip
\noindent\rmfamily\textbf{Laser linewidth measurement:}
The intrinsic linewidth of the laser is characterized by a variation of the self-heterodyne measurement \cite{yuan2022correlated}. A splitter separates the light into two paths in a Mach-Zehnder interferometer (MZI), with a 17-m delay line in one arm and an AOM in another. The AOM is modulated at 300 MHz, creating an isolation of the low-frequency environment noise from the measurement. The laser waves output from the MZI are then detected by a balanced photodetector (BPD) whose output is recorded by a real-time oscilloscope. The phase noise is extracted from the captured waveform via the Hilbert transform and converted into the frequency noise from which we can identify the white noise floor and the corresponding intrinsic laser linewidth. Additionally, we employ a second BPD in the setup to remove the BPD introduced noise in the system using the cross-correlation technique. {Fig.~\ref{Fig2}(c)} shows the derived frequency noises for device 1 and 2 with the BPD noise cancellation. Compared to the results without BPD noise cancellation, a clear noise reduction can be observed at the offset frequency larger than 5 MHz resulting in about 36 Hz lower measured intrinsic linewidth of our laser for both device 1 and device 2. A more detailed analysis is given in the SI.

\def\bibsection{\section*{\refname}} 
\bibliography{all_reference.bib}

\bibliographystyle{naturemag}

\if{

}\fi
 
    \medskip
	\newpage
	\noindent\sffamily\textbf{Acknowledgements}
	
	\noindent\rmfamily The authors thank Joel Guo and Linrui Tan for valuable discussions and help on experiment. This work is supported in part by the Defense Advanced Research Projects Agency (DARPA) LUMOS program under Agreement No.~HR001-20-2-0044, and the National Science Foundation (NSF) (OMA-2138174, ECCS-2231036, OSI-2329017). This work was performed in part at the Cornell NanoScale Facility, a member of the National Nanotechnology Coordinated Infrastructure (National Science Foundation, ECCS-1542081); at the Cornell Center for Materials Research (National Science Foundation, Grant No. DMR-1719875); and in the UCSB Nanofabrication Facility, an open access laboratory.\newline

	\noindent\textbf{Author Contributions} S.X., M.L., and Q.L. conceived the experiment. S.X. and M.L. performed numerical simulations. S.X. and M.L. fabricated the device. S.X., M.L. carried out the device characterization. J.L., Z.G., Q.H.,  R.L., T.Q. and J.S. provided helpful suggestions in data measurement and processing. H.W. helped on the characterization of laser linewidth. L.C. and C.X. provided valuable suggestions to the design of device. S.X. and M.L. wrote the manuscript with contributions from all authors. Q.L. and J.B. supervised the project. Q.L. conceived the device concept. \newline


	
	\noindent\textbf{Competing interests.} The authors declare no competing interests.  \newline
	
	\noindent\textbf{Data and materials availability} The data used to produce the plots within this work will be released upon publication of this preprint.

\end{document}